\documentclass[twocolumn,draft,showkeys,showpacs,eqsecnum,nofootinbib,aps]{revtex4}
\renewcommand{\theequation}{\arabic{equation}}
\def\beq{\begin{equation}}
\def\eeq{\end{equation}}
\def\bea{\begin{eqnarray}}
\def\eea{\end{eqnarray}}\def\nn{\nonumber}

\begin{document}
\title{Sum rules for baryon decuplet magnetic moments}
\author{Soon-Tae Hong}
\email{soonhong@ewha.ac.kr} \affiliation{W.K. Kellogg Radiation
Laboratory, California Institute of Technology, Pasadena, CA 91125
USA} \affiliation{Department of Science Education and Research
Institute for Basic Sciences, Ewha Womans University, Seoul
120-750 Korea}

\date{\today}
\begin{abstract}
In chiral models with SU(3) group structure, baryon decuplet and
octet magnetic moments are evaluated by constructing their sum
rules to yield theoretical predictions.  In these sum rules we
exploit six experimentally known baryon magnetic moments.  Sum
rules for flavor components and strange form factors of the octet
and decuplet magnetic moments and decuplet-to-octet transition
magnetic moments are also investigated.
\end{abstract}
\pacs{12.39.-x, 11.30.Hv, 21.60.Fw, 13.40.Em, 13.40.Gp, 14.20.-c}
\keywords{sum rules, baryon decuplet and octet, magnetic moments,
strange form factors, chiral model} \maketitle

\section{Introduction}
\setcounter{equation}{0}
\renewcommand{\theequation}{\arabic{section}.\arabic{equation}}

The internal structure of the nucleon is still a subject of great
interest to both experimentalists and theorists. In 1933, Frisch
and Stern~\cite{stern33} performed the first measurement of the
magnetic moment of the proton and obtained the earliest
experimental evidence for the internal structure of the nucleon.
However, it wasn't until forty years later that the quark
structure of the nucleon was directly observed in deep inelastic
electron scattering experiments and we still lack a quantitative
theoretical understanding of these properties including the
magnetic moments.

Since Coleman and Glashow~\cite{cg} predicted the magnetic moments
of the baryon octet about forty years ago, there has been a lot of
progress in both the theoretical paradigm and experimental
verification for the baryon magnetic moments. The measurements of
the baryon decuplet magnetic moments were reported for
$\mu_{\Delta^{++}}$~\cite{boss} and $\mu_{\Omega^{-}}$~\cite{die}
to yield a new avenue for understanding the hadron structure. The
magnetic moments of baryon decuplet have been theoretically
investigated in several models such as the quenched lattice gauge
theory~\cite{lein}, the quark models~\cite{sch}, the chiral bag
model~\cite{decup}, the chiral perturbation theory~\cite{but}, the
QCD sum rules~\cite{lee1}, the chiral quark model~\cite{lin2} and
chiral quark soliton model~\cite{kim04}. Moreover, by including
the effect of decuplet intermediate states of spin-3/2 baryons
explicitly, the heavy baryon chiral expansion of baryon octet
magnetic moments~\cite{jenkins93plb} and charge
radii~\cite{musolf01} were investigated. The decuplet-to-octet
transition magnetic moments have been also analyzed in the
$1/N_{c}$ expansion of
QCD~\cite{jenkins94plb,jenkins02prl,lebed04} and in the chiral
quark soliton model~\cite{kim05}.

Quite recently, the SAMPLE Collaboration~\cite{sample04} reported
the experimental data of the proton strange form factor through
parity violating electron scattering~\cite{mck89}.  To be more
precise, they measured the neutral weak form factors at a small
momentum transfer $Q_{S}^2 = 0.1~{\rm (GeV/c)}^2$ to yield the
proton strange magnetic form factor in units of Bohr nuclear
magnetons (n.m.) $G_{M}^{s}=+0.37 \pm 0.20\pm 0.26 \pm 0.07$
n.m.~\cite{sample04}.  The HAPPEX Collaboration later reported
$G_{M}^{s}=+0.18\pm 0.27$ n.m.~\cite{happex06}.  Moreover,
McKeown~\cite{mck} has shown that the strange form factor of
proton should be positive by using the conjecture that the
up-quark effects are generally dominant in the flavor dependence
of the nucleon properties.  The chiral bag model~\cite{gerry791}
predicted firstly the positive value for the proton strange form
factor~\cite{hong93}.

In this paper, we will exploit the chiral bag model to predict
baryon decuplet and octet magnetic moments and their strange form
factors. This model calculation can share those of other Skyrmion
extended models with SU(3) group structure if sum rules are
properly used. More specifically, in the chiral models, we will
investigate the magnetic moments of baryon decuplet and octet,
together with the decuplet-to-octet transition magnetic moments,
in terms of their sum rules. We will also study the sum rules for
the flavor components and strange form factors of the baryon
magnetic moments. In Section 2, in the adjoint representation we
construct the sum rules of the baryon decuplet and octet magnetic
moments and the transition magnetic moments in the SU(3) chiral
models, and in Section 3 the sum rules for their flavor components
and strange form factors in a model independent way at least in
the category of the chiral models with the SU(3) flavor group. In
Appendix A, we list the u- and d-flavor components of the baryon
magnetic moments and transition magnetic moments and their sum
rules.

\section{Magnetic moments of baryon decuplet}
\setcounter{equation}{0}
\renewcommand{\theequation}{\arabic{section}.\arabic{equation}}

We start with the chiral bag model with the broken U-spin symmetry
whose Lagrangian is of the form \bea
&&{\cal L}={\cal L}_{CS}+{\cal L}_{CSB}+{\cal L}_{FSB}\nn\\
&&{\cal L}_{CS}=\bar{\psi}i\gamma^{\mu}\partial_{\mu}\psi\Theta_{B} -\frac{1}{2}\bar{%
\psi}U_{5}\psi\Delta_{B}\nn\\
&&~~+\left(-\frac{1}{4}f_{\pi}^{2}{\rm tr}(l_{\mu}l^{\mu}) +%
\frac{1}{32e^{2}}[l_{\mu},l_{\nu}]^{2}+{\cal
L}_{WZW}\right)\bar{\Theta}_{B}
\nonumber \\
&&{\cal L}_{CSB}=-\bar{\psi}M\psi\Theta_{B}+\frac{1}{4}f_{\pi}^{2}m_{%
\pi}^{2} {\rm tr}(U+U^{\dagger}-2)\bar{\Theta}_{B}  \nonumber \\
&&{\cal L}_{FSB}=\frac{1}{6}f_{\pi}^{2}(\chi^{2}m_{K}^{2}-m_{\pi}^{2})\nn\\
&&~~\cdot{\rm
tr}((1-\sqrt{3}\lambda_{8})(U+U^{\dagger}-2))\bar{\Theta}_{B}
\nn\\
&&~~-\frac{1}{12}f_{\pi}^{2}(\chi^{2}-1){\rm tr}((1-\sqrt{3}
\lambda_{8})(Ul_{\mu}l^{\mu}+l_{\mu}l^{\mu}U^{\dagger}))\bar{\Theta}_{B}\nn\\
\label{lag} \eea where the quark field $\psi$ has SU(3) flavor
degrees of freedom and the chiral field
$U=e^{i\lambda_{a}\pi_{a}/f_{\pi}}\in$ SU(3) is described by
the pseudoscalar meson fields $\pi_{a}$ (a=1,...8) and Gell-Mann matrices $%
\lambda_{a}$ with
$\lambda_{a}\lambda_{b}=\frac{2}{3}\delta_{ab}+(if_{abc}
+d_{abc})\lambda_{c}$, and $\Theta_{B}(=1-\bar{\Theta}_{B})$ is
the bag theta function (one inside the bag and zero outside the
bag). In the limit of vanishing bag radius, the chiral bag model
is reduced to the Skyrmion model. Here
$l_{\mu}=U^{\dagger}\partial_{\mu}U$ and ${\cal L}_{WZW}$ stands
for the topological Wess-Zumino-Witten (WZW) term. The chiral
symmetry (CS) is broken by the quark masses $M={\rm diag}
(m_{u},m_{d},m_{s})$ and pion mass $m_{\pi}$ in ${\cal L}_{CSB}$.
Furthermore the SU(3) flavor symmetry breaking (FSB) with $m_{K}/m_{\pi}\neq 1$ and $%
\chi=f_{K}/f_{\pi}\neq 1$ is included in ${\cal L}_{FSB}$.  Even
though the mass terms in ${\cal L}_{CSB}$ and ${\cal L}_{FSB}$
break both the SU$_{L}$(3)$\times$SU$_{R}$(3) and diagonal SU(3)
symmetry so that chiral symmetry cannot be conserved, these terms
without derivatives yield no explicit contribution to the
electromagnetic (EM) currents $J^{\mu}$ and at least in the
adjoint representation of the SU(3) group the EM currents are
conserved and of the same form as the chiral limit result
$J^{\mu}_{CS}$
to preserve the U-spin symmetry. However the derivative-dependent term in $%
{\cal L}_{FSB}$ gives rise to the U-spin symmetry breaking conserved EM
currents $J^{\mu}_{FSB}$ so that $J^{\mu}=J^{\mu}_{CS}+J^{\mu}_{FSB}$.

Assuming that the hedgehog classical solution in the meson phase
$U_{0}= e^{i\lambda_{i}\hat{r}_{i}\theta (r)}$ (i=1,2,3) is
embedded in the SU(2) isospin subgroup of SU(3) and the Fock space
in the quark phase is described by the $N_{c}$ valence quarks and
the vacuum structure composed of quarks filling the negative
energy sea, the chiral model generates the zero mode with the
collective variable $A(t)\in$ SU(3) by performing the slow rotation $%
U\rightarrow AU_{0} A^{\dagger}$ and $\psi\rightarrow A\psi$ on
SU(3) group manifold.  Given the spinning chiral model ansatz, the
EM currents yield the magnetic moment operators
$\hat{\mu}^{i}=\hat{\mu}^{i(3)}+\frac{1}{\sqrt{3}}\hat{\mu}^{i(8)}$
where
$\hat{\mu}^{i(a)}=\hat{\mu}^{i(a)}_{CS}+\hat{\mu}^{i(a)}_{FSB}$
with
\begin{eqnarray}
\hat{\mu}^{i(a)}_{CS}&=&-{\cal N}D_{ai}^{8}-{\cal N}^{%
\prime}d_{ipq}D_{ap}^{8} \hat{T}_{q}^{R}+\frac{N_{c}}{2\sqrt{3}}{\cal M}%
D_{a8}^{8}\hat{J}_{i}  \nonumber \\
\hat{\mu}^{i(a)}_{FSB}&=&-{\cal P}D_{ai}^{8}(1-D_{88}^{8})+{\cal Q} \frac{%
\sqrt{3}}{2}d_{ipq}D_{ap}^{8}D_{8q}^{8}\nn\\
& &+{\cal R}D_{a8}^{8}D_{8i}^{8} \label{magop}
\end{eqnarray}
where ${\cal M}$, ${\cal N}$, ${\cal N}^{\prime}$, ${\cal P}$,
${\cal Q}$ and ${\cal R}$ are the inertia parameters calculable in
the chiral models~\cite{hong93,hongpr01}.  Using the theorem that
the tensor product of the Wigner D functions can be decomposed
into sum of the single D functions, the isovector and isoscalar
parts of the operator $\hat{\mu}^{i(a)}_{FSB}$ are then rewritten
as
\begin{eqnarray}
\hat{\mu}^{i(3)}_{FSB}&=&{\cal
P}\left(-\frac{4}{5}D_{3i}^{8}+\frac{1}{4}D_{3i}^{10}+\frac{1}{4}D_{3i}^{\bar{10}}
+\frac{3}{10}D_{3i}^{27}\right)\nn\\
& &+{\cal Q
}\left(\frac{3}{10}D_{3i}^{8}-\frac{3}{10}D_{3i}^{27}\right)\nn\\
& &+{\cal R}\left(\frac{1}{5}D_{3i}^{8}+\frac{7}{10}D_{3i}^{27}\right)\nonumber \\
\hat{\mu}^{i(8)}_{FSB}&=&{\cal
P}\left(-\frac{6}{5}D_{8i}^{8}+\frac{9}{20}
D_{8i}^{27}\right)\nn\\
& &+{\cal
Q}\left(-\frac{3}{10}D_{8i}^{8}-\frac{9}{20}D_{8i}^{27}\right)
\nn\\
& &+{\cal
R}\left(-\frac{1}{5}D_{8i}^{8}+\frac{9}{20}D_{8i}^{27}\right).
\end{eqnarray}
Here one notes that the ${\bf 1}$, ${\bf 10}$ and $\bar{{\bf 10}}$
irreducible representations (IRs) do not occur in the decuplet baryons while
${\bf 10}$ and $\bar{{\bf 10}}$ IRs appear together in the isovector channel
of the baryon octet to conserve the hermiticity of the operator.

Using the above operator $\hat{\mu}^{i}$ together with the
decuplet baryon wave function $\Phi_{B}^{\lambda}= \sqrt{%
{\rm dim}(\lambda)}D_{ab}^{\lambda}$ with the quantum numbers
$a=(Y;I,I_{3})$ ($Y$; hypercharge, $I$; isospin) and
$b=(Y_{R};J,-J_{3})$ ($Y_{R}$; right hypercharge, $J$; spin) and
$\lambda$ the dimension of the representation, the baryon decuplet
magnetic moments for ${\bf 10}~(J_{3}=3/2)$ and transition
magnetic moments for ${\bf 10}~ (J_{3}=1/2) \rightarrow {\bf
8}~(J_{3}=1/2)+\gamma$ have the following hyperfine structure
\begin{eqnarray}
\mu_{\Delta^{++}}&=&\frac{1}{8}{\cal M}+\frac{1}{2}({\cal N}-\frac{1} {2%
\sqrt{3}}{\cal N}^{\prime})+\frac{3}{7}{\cal P}\nn\\
& &-\frac{3}{56}{\cal Q}-\frac{1}{14}{\cal R}
\nonumber \\
\mu_{\Delta^{+}}&=&\frac{1}{16}{\cal M}+\frac{1}{4}({\cal N}-\frac{1} {2%
\sqrt{3}}{\cal N}^{\prime})+\frac{5}{21}{\cal P}\nn\\
&&+\frac{1}{84}{\cal Q}-\frac{1}{84}{\cal R}
\nonumber \\
\mu_{\Delta^{0}}&=&\frac{1}{21}{\cal P}+\frac{13}{168}{\cal Q}+\frac{1}{21}{\cal R}  \nonumber \\
\mu_{\Delta^{-}}&=&-\frac{1}{16}{\cal M}-\frac{1}{4}({\cal N}-\frac{1} {2%
\sqrt{3}}{\cal N}^{\prime})-\frac{1}{7}{\cal P}\nn\\
&&+\frac{1}{7}{\cal Q}+\frac{3}{28}{\cal R}
\nonumber \\
\mu_{\Sigma^{*+}}&=&\frac{1}{16}{\cal M}+\frac{1}{4}({\cal N}-\frac{1} {2%
\sqrt{3}}{\cal N}^{\prime})+\frac{19}{84}{\cal
P}\nn\\
&&-\frac{17}{168}{\cal Q}-\frac{1}{42}{\cal R}
\nonumber \\
\mu_{\Sigma^{*0}}&=&\frac{1}{84}{\cal P}-\frac{1}{84}{\cal Q}+\frac{1}{84}{\cal R}  \nonumber \\
\mu_{\Sigma^{*-}}&=&-\frac{1}{16}{\cal M}-\frac{1}{4}({\cal N}-\frac{1} {2%
\sqrt{3}}{\cal N}^{\prime})-\frac{17}{84}{\cal
P}\nn\\
&&+\frac{13}{168}{\cal Q}+\frac{1}{21}{\cal R}
\nonumber \\
\mu_{\Xi^{*0}}&=&-\frac{1}{42}{\cal P}-\frac{17}{168}{\cal Q}-\frac{1}{42}{\cal R}  \nonumber \\
\mu_{\Xi^{*-}}&=&-\frac{1}{16}{\cal M}-\frac{1}{4}({\cal N}-\frac{1} {2\sqrt{%
3}}{\cal N}^{\prime})-\frac{11}{42}{\cal P}\nn\\
&&+\frac{1}{84}{\cal Q}-\frac{1}{84}{\cal R}  \nonumber
\\
\mu_{\Omega^{-}}&=&-\frac{1}{16}{\cal M}-\frac{1}{4}({\cal N}-\frac{1} {2%
\sqrt{3}}{\cal N}^{\prime})-\frac{9}{28}{\cal P}\nn\\
&&-\frac{3}{56}{\cal
Q}-\frac{1}{14}{\cal R}\nn\\
\frac{1}{\sqrt{5}}\mu_{p\Delta^{+}}&=&\frac{1}{\sqrt{5}}\mu_{n\Delta^{0}}\nn\\
&=&-\frac{2}{15}({\cal N}+\frac{2-\sqrt{2}}{8}{\cal
N}^{\prime})-\frac{4}{45}{\cal
P}\nn\\
&&+\frac{7}{180}{\cal
Q}+\frac{79}{2700}{\cal R}\nn\\
\frac{1}{\sqrt{5}}\mu_{\Sigma^{+}\Sigma^{*+}}&=&\frac{2}{15}({\cal
N}+\frac{2-\sqrt{2}}{8}{\cal N}^{\prime})+\frac{13}{90}{\cal
P}\nn\\
&&+\frac{1}{180}{\cal
Q}-\frac{1}{100}{\cal R}\nn\\
\frac{1}{\sqrt{5}}\mu_{\Sigma^{0}\Sigma^{*0}}&=&\frac{1}{15}({\cal
N}+\frac{2-\sqrt{2}}{8}{\cal N}^{\prime})+\frac{7}{90}{\cal
P}\nn\\
&&+\frac{1}{45}{\cal Q}+\frac{1}{90}{\cal R}\nn\\
\frac{1}{\sqrt{5}}\mu_{\Sigma^{-}\Sigma^{*-}}&=&\frac{1}{90}{\cal
P}+\frac{7}{180}{\cal Q}+\frac{29}{900}{\cal R}\nn\\
\frac{1}{\sqrt{5}}\mu_{\Xi^{0}\Xi^{*0}}&=&\frac{2}{15}({\cal
N}+\frac{2-\sqrt{2}}{8}{\cal N}^{\prime})+\frac{7}{45}{\cal
P}\nn\\
&&-\frac{1}{180}{\cal Q}+\frac{23}{2700}{\cal R}\nn\\
\frac{1}{\sqrt{5}}\mu_{\Xi^{-}\Xi^{*-}}&=&\frac{1}{90}{\cal P}
+\frac{7}{180}{\cal Q}+\frac{67}{2700}{\cal R}\nn\\
\frac{1}{\sqrt{15}}\mu_{\Lambda\Sigma^{*0}}&=&-\frac{1}{15}({\cal
N}+\frac{2-\sqrt{2}}{8}{\cal N}^{\prime})-\frac{1}{18}{\cal
P}\nn\\
&&+\frac{1}{45}{\cal Q}+\frac{11}{1350}{\cal R}.\label{dec}
\end{eqnarray}
Here the coefficients are solely given by the SU(3) group
structure of the chiral models and the physical informations such
as decay constants and masses are included in the above inertia
parameters, such as ${\cal M}$, ${\cal N}$ and so on.  Note that
the SU(3) group structure in the coefficients is generic property
shared by the chiral models which exploit the hedgehog ansatz
solution corresponding to the little group SU(2)$\times {\bf
Z}_{2}$~\cite{jenkins94}. In the chiral perturbation
theory~\cite{but} and in the $1/N_{c}$ expansion of
QCD~\cite{jenkins94plb,jenkins02prl,lebed04} to which the hedgehog
ansatz does not apply, one can thus see the coefficients different
from those in (\ref{dec}) even though the SU(3) flavor group is
used in the theory. Now, it seems appropriate to comment on the
$1/N_{c}$ expansion~\cite{thooft74,witten79,jenkins94,jenkins98}.
In the above relations (\ref{dec}), the inertia parameters ${\cal
N}$, ${\cal N}^{\prime}$, ${\cal P}$, ${\cal Q}$ and ${\cal R}$
are of order $N_{c}$ while ${\cal M}$ is of order $N_{c}^{-1}$.
However, since the inertia parameter ${\cal M}$ is multiplied by
an explicit factor $N_{c}$ in (\ref{magop}), the terms with ${\cal
M}$ are of order $N_{c}^{0}$.  (For details of further $1/N_{c}$,
see~\cite{jenkins94,jenkins98}.)

In the SU(3) flavor symmetric limit with the chiral symmetry
breaking masses $m_{u}=m_{d}=m_{s}$, $m_{K}=m_{\pi}$ and decay
constants $f_{K}=f_{\pi}$, the magnetic moments of the decuplet
baryons are simply given by~\cite{lee}
\begin{equation}
\mu_{B}=Q_{EM}\left(\frac{1}{16}{\cal M}+\frac{1}{4}\left({\cal
N}-\frac{1}{2\sqrt{3}} {\cal N}^{\prime}\right)\right)  \label{q}
\end{equation}
where $Q_{EM}$ is the EM charge. Here one notes that in the chiral
model in the adjoint representation the prediction of the baryon
magnetic moments with the chiral symmetry is the same as that with
the SU(3) flavor symmetry since the mass-dependent term in ${\cal
L}_{CSB}$ and ${\cal L}_{FSB}$ do not
yield any contribution to $J^{\mu}_{FSB}$ so that there is no terms with $%
{\cal P}$, ${\cal Q}$ and ${\cal R}$ in (\ref{dec}). Due to the
degenerate d- and s-flavor charges in the SU(3) EM charge operator
$\hat{Q}_{EM}$, the chiral model possesses the U-spin symmetry
relations in the baryon decuplet magnetic moments, similar to
those in the octet baryons~\cite{nrqm}
\begin{eqnarray}
\mu_{\Delta^{-}}&=&\mu_{\Sigma^{*-}}=\mu_{\Xi^{*-}}=\mu_{\Omega^{-}}
\nonumber \\
\mu_{\Delta^{0}}&=&\mu_{\Sigma^{*0}}=\mu_{\Xi^{*0}}  \nonumber \\
\mu_{\Delta^{+}}&=&\mu_{\Sigma^{*+}}  \label{uspin}
\end{eqnarray}
which are subset of the more strong symmetry relations (\ref{q}).
Next, since the SU(3) FSB quark masses do not affect the magnetic
moments of the baryon decuplet in the adjoint representation of
the chiral model, in the more
general SU(3) flavor symmetry broken case with $m_{u}=m_{d}\neq m_{s}$, $%
m_{\pi}\neq m_{K}$ and $f_{\pi}\neq f_{K}$, the decuplet baryon
magnetic moments with all the inertia parameters satisfy the
following symmetric sum rules
\begin{eqnarray}
\mu_{\Sigma^{*0}}&=&\frac{1}{2}\mu_{\Sigma^{*+}}+\frac{1}{2}\mu_{\Sigma^{*-}}
\nn\\
\mu_{\Delta^{-}}+\mu_{\Delta^{++}}&=&\mu_{\Delta^{0}}+\mu_{\Delta^{+}}
\nn\\
\sum_{B\in {\rm decuplet}}\mu_{B}&=&0.  \label{cg}
\end{eqnarray}

Now, in order to predict baryon decuplet magnetic moments we
proceed to derive sum rules for the baryon magnetic moments in
terms of the experimentally known baryon magnetic moments.  To do
this, we first consider the baryon octet magnetic moments of the
form
\bea \mu_{p}&=&\frac{1}{10}{\cal M}+\frac{4}{15}({\cal
N}+\frac{1}{2} {\cal N}^{\prime})+\frac{8}{45}{\cal
P}\nn\\
& &-\frac{2}{45}{\cal Q}-\frac{1}{45}{\cal R}
\nonumber \\
\mu_{n}&=&\frac{1}{20}{\cal M}-\frac{1}{5}({\cal N}+\frac{1}{2}
{\cal N}^{\prime})-\frac{1}{9}{\cal P}\nn\\
& &+\frac{7}{90}{\cal Q}+\frac{1}{45}{\cal R}
\nonumber \\
\mu_{\Sigma^{+}}&=&\frac{1}{10}{\cal M}+\frac{4}{15}({\cal
N}+\frac{1}{2} {\cal N}^{\prime})+\frac{13}{45}{\cal
P}\nn\\
& &-\frac{1}{45}{\cal Q}-\frac{2}{45}{\cal R}
\nonumber \\
\mu_{\Sigma^{0}}&=&-\frac{1}{40}{\cal M}+\frac{1}{10}({\cal
N}+\frac{1}{2} {\cal N}^{\prime})+\frac{11}{90}{\cal
P}\nn\\
& &+\frac{1}{36}{\cal Q}+\frac{1}{45}{\cal R}
\nonumber \\
\mu_{\Sigma^{-}}&=&-\frac{3}{20}{\cal M}-\frac{1}{15}({\cal
N}+\frac{1}{2} {\cal N}^{\prime})-\frac{2}{45}{\cal
P}\nn\\
& &+\frac{7}{90}{\cal Q}+\frac{4}{45}{\cal R} \nonumber\\
\mu_{\Xi^{0}}&=&\frac{1}{20}{\cal M}-\frac{1}{5}({\cal
N}+\frac{1}{2} {\cal N}^{\prime})-\frac{11}{45}{\cal
P}\nn\\
& &-\frac{1}{45}{\cal Q}+\frac{1}{45}{\cal R}
\nonumber \\
\mu_{\Xi^{-}}&=&-\frac{3}{20}{\cal M}-\frac{1}{15}({\cal
N}+\frac{1}{2} {\cal N}^{\prime})-\frac{4}{45}{\cal
P}\nn\\
& &-\frac{2}{45}{\cal Q}-\frac{4}{45}{\cal R}
\nonumber \\
\mu_{\Lambda}&=&\frac{1}{40}{\cal M}-\frac{1}{10}({\cal
N}+\frac{1}{2} {\cal N}^{\prime})-\frac{1}{10}{\cal
P}\nn\\
& &-\frac{1}{20}{\cal Q}. \label{octet} \eea Since we have
effectively five inertia parameters ${\cal M}$, ${\cal
N}+\frac{1}{2}{\cal N}^{\prime}$, ${\cal P}$, ${\cal Q}$ and
${\cal R}$, we can derive sum rules for three magnetic moments
$\mu_{\Sigma^{0}}$, $\mu_{\Lambda}$ and $\mu_{\Xi^{-}}$ in terms
of five experimentally known octet magnetic moments, $\mu_{p}$,
$\mu_{n}$, $\mu_{\Sigma^{+}}$, $\mu_{\Sigma^{-}}$ and
$\mu_{\Xi^{0}}$ as follows \bea
\mu_{\Sigma^{0}}&=&\frac{1}{2}\mu_{\Sigma^{+}}+\frac{1}{2}\mu_{\Sigma^{-}}\nonumber\\
\mu_{\Xi^{-}}&=&-\frac{1}{3}\mu_{n}-\frac{8}{3}\mu_{\Sigma^{+}}
-\frac{5}{3}\mu_{\Sigma^{-}}-\frac{7}{3}\mu_{\Xi^{0}}\nonumber\\
\mu_{\Lambda}&=&-\mu_{p}-\frac{2}{3}\mu_{n}+\frac{7}{6}\mu_{\Sigma^{+}}
+\frac{1}{6}\mu_{\Sigma^{-}}+\frac{4}{3}\mu_{\Xi^{0}}.
\label{octetsum} \eea Using the above sum ruels we can predict the
magnetic moments as in Table 1.  One notes that the value of
$\mu_{\Lambda}$ is comparable to the experimental data
$\mu_{\Lambda}^{exp}=-0.61$, while the value of $\mu_{\Xi^{-}}$ is
not so comparable to $\mu_{\Xi^{-}}^{exp}=-0.65$.

Similarly we can derive sum rules for the decuplet magnetic
moments in terms of six experimentally known magnetic moments,
$\mu_{p}$, $\mu_{n}$, $\mu_{\Sigma^{+}}$, $\mu_{\Sigma^{-}}$,
$\mu_{\Xi^{0}}$ and $\mu_{\Delta^{++}}$ to arrive at
\begin{widetext}\begin{eqnarray}
\mu_{\Delta^{+}}&=&\frac{5}{28}\mu_{p}+\frac{55}{168}\mu_{n}+\frac{5}{42}
\mu_{\Sigma^{+}}+\frac{25}{84}\mu_{\Sigma^{-}}
-\frac{5}{168}\mu_{\Xi^{0}}+\frac{1}{2}\mu_{\Delta^{++}}
\nonumber \\
\mu_{\Delta^{0}}&=&\frac{5}{14}\mu_{p}+\frac{55}{84}\mu_{n}+\frac{5}{21}
\mu_{\Sigma^{+}}+\frac{25}{42}\mu_{\Sigma^{-}}-\frac{5}{84}\mu_{\Xi^{0}}
\nonumber \\
\mu_{\Delta^{-}}&=&\frac{15}{28}\mu_{p}+\frac{55}{56}\mu_{n}+\frac{5}{14}
\mu_{\Sigma^{+}}+\frac{25}{28}\mu_{\Sigma^{-}}
-\frac{5}{56}\mu_{\Xi^{0}}-\frac{1}{2}\mu_{\Delta^{++}}\nonumber \\
\mu_{\Sigma^{*+}}&=&-\frac{25}{14}\mu_{p}-\frac{235}{168}\mu_{n}+\frac{215}{84}
\mu_{\Sigma^{+}}+\frac{65}{84}\mu_{\Sigma^{-}}
+\frac{365}{168}\mu_{\Xi^{0}}+\frac{1}{2}\mu_{\Delta^{++}}
\nonumber \\
\mu_{\Sigma^{*0}}&=&-\frac{15}{28}\mu_{p}-\frac{5}{14}\mu_{n}+\frac{25}{28}
\mu_{\Sigma^{+}}+\frac{5}{14}\mu_{\Sigma^{-}}+\frac{5}{7}\mu_{\Xi^{0}}
\nonumber \\
\mu_{\Sigma^{*-}}&=&\frac{5}{7}\mu_{p}+\frac{115}{168}\mu_{n}-\frac{65}{84}
\mu_{\Sigma^{+}}-\frac{5}{84}\mu_{\Sigma^{-}}
-\frac{125}{168}\mu_{\Xi^{0}}-\frac{1}{2}\mu_{\Delta^{++}}\nonumber \\
\mu_{\Xi^{*0}}&=&-\frac{10}{7}\mu_{p}-\frac{115}{84}\mu_{n}+\frac{65}{42}
\mu_{\Sigma^{+}}+\frac{5}{42}\mu_{\Sigma^{-}}+\frac{125}{84}\mu_{\Xi^{0}}
\nonumber \\
\mu_{\Xi^{*-}}&=&\frac{25}{28}\mu_{p}+\frac{65}{168}\mu_{n}-\frac{40}{21}
\mu_{\Sigma^{+}}-\frac{85}{84}\mu_{\Sigma^{-}}
-\frac{235}{168}\mu_{\Xi^{0}}-\frac{1}{2}\mu_{\Delta^{++}}\nonumber\\
\mu_{\Omega^{-}}&=&\frac{15}{14}\mu_{p}+\frac{5}{56}\mu_{n}-\frac{85}{28}
\mu_{\Sigma^{+}}-\frac{55}{28}\mu_{\Sigma^{-}}
-\frac{115}{56}\mu_{\Xi^{0}}-\frac{1}{2}\mu_{\Delta^{++}}\nn\\
\frac{1}{\sqrt{5}}\mu_{p\Delta^{+}}&=&
\frac{1}{\sqrt{5}}\mu_{n\Delta^{0}}\nonumber\\
&=&\left(-\frac{1}{189}+\frac{17\sqrt{2}}{140}
-\frac{17\sqrt{3}}{210}-\frac{17\sqrt{6}}{420}\right)\mu_{p}
+\left(\frac{173}{567}+\frac{\sqrt{2}}{168}
-\frac{\sqrt{3}}{252}-\frac{\sqrt{6}}{504}\right)\mu_{n}\nonumber\\
& &+\left(\frac{3383}{11340}-\frac{\sqrt{2}}{420}
+\frac{\sqrt{3}}{630}+\frac{\sqrt{6}}{1260}\right)\mu_{\Sigma^{+}}
+\left(\frac{2189}{11340}-\frac{13\sqrt{2}}{420}+\frac{13\sqrt{3}}{630}
+\frac{13\sqrt{6}}{1260}\right)\mu_{\Sigma^{-}}\nonumber\\
& &+\left(\frac{2131}{11340}-\frac{73\sqrt{2}}{840}
+\frac{73\sqrt{3}}{1260}+\frac{73\sqrt{6}}{2520}\right)\mu_{\Xi^{0}}
+\left(-\frac{1}{5}-\frac{\sqrt{2}}{10}
+\frac{\sqrt{3}}{15}+\frac{\sqrt{6}}{30}\right)\mu_{\Delta^{++}}\nonumber\\
\frac{1}{\sqrt{5}}\mu_{\Sigma^{+}\Sigma^{*+}}
&=&\left(-\frac{83}{630}-\frac{17\sqrt{2}}{140}
+\frac{17\sqrt{3}}{210}+\frac{17\sqrt{6}}{420}\right)\mu_{p}
+\left(-\frac{1}{378}-\frac{\sqrt{2}}{168}
+\frac{\sqrt{3}}{252}+\frac{\sqrt{6}}{504}\right)\mu_{n}\nonumber\\
& &+\left(\frac{131}{756}+\frac{\sqrt{2}}{420}
-\frac{\sqrt{3}}{630}-\frac{\sqrt{6}}{1260}\right)\mu_{\Sigma^{+}}
+\left(\frac{107}{756}
+\frac{13\sqrt{2}}{420}-\frac{13\sqrt{3}}{630}-\frac{13\sqrt{6}}{1260}\right)\mu_{\Sigma^{-}}\nonumber\\
&&+\left(-\frac{589}{3780}+\frac{73\sqrt{2}}{840}
-\frac{73\sqrt{3}}{1260}-\frac{73\sqrt{6}}{2520}\right)\mu_{\Xi^{0}}
+\left(\frac{1}{5}+\frac{\sqrt{2}}{10}
-\frac{\sqrt{3}}{15}-\frac{\sqrt{6}}{30}\right)\mu_{\Delta^{++}}\nonumber\\
\frac{1}{\sqrt{5}}\mu_{\Sigma^{0}\Sigma^{*0}}&=&\left(\frac{11}{140}-\frac{17\sqrt{2}}{280}
+\frac{17\sqrt{3}}{420}+\frac{17\sqrt{6}}{840}\right)\mu_{p}
+\left(\frac{9}{56}-\frac{\sqrt{2}}{336}
+\frac{\sqrt{3}}{504}+\frac{\sqrt{6}}{1008}\right)\mu_{n}\nonumber\\
& &+\left(\frac{19}{140}+\frac{\sqrt{2}}{840}
-\frac{\sqrt{3}}{1260}-\frac{\sqrt{6}}{2520}\right)\mu_{\Sigma^{+}}
+\left(\frac{37}{140}
+\frac{13\sqrt{2}}{840}-\frac{13\sqrt{3}}{1260}-\frac{13\sqrt{6}}{2520}\right)\mu_{\Sigma^{-}}\nonumber\\
&&+\left(-\frac{13}{280}+\frac{73\sqrt{2}}{1680}
-\frac{73\sqrt{3}}{2520}-\frac{73\sqrt{6}}{5040}\right)\mu_{\Xi^{0}}
+\left(\frac{1}{10}+\frac{\sqrt{2}}{20}
-\frac{\sqrt{3}}{30}-\frac{\sqrt{6}}{60}\right)\mu_{\Delta^{++}}\nonumber\\
\frac{1}{\sqrt{5}}\mu_{\Sigma^{-}\Sigma^{*-}}&=&
\frac{13}{45}\mu_{p}+\frac{35}{108}\mu_{n}+\frac{53}{540}\mu_{\Sigma^{+}}
+\frac{209}{540}\mu_{\Sigma^{-}}+\frac{17}{270}\mu_{\Xi^{0}}\nonumber\\
\frac{1}{\sqrt{5}}\mu_{\Xi^{0}\Xi^{*0}}&=&\left(-\frac{667}{945}-\frac{17\sqrt{2}}{140}
+\frac{17\sqrt{3}}{210}+\frac{17\sqrt{6}}{420}\right)\mu_{p}
+\left(-\frac{451}{1134}-\frac{\sqrt{2}}{168}
+\frac{\sqrt{3}}{252}+\frac{\sqrt{6}}{504}\right)\mu_{n}\nonumber\\
& &+\left(\frac{2759}{2268}+\frac{\sqrt{2}}{420}
-\frac{\sqrt{3}}{630}-\frac{\sqrt{6}}{1260}\right)\mu_{\Sigma^{+}}
+\left(\frac{1385}{2268}+\frac{13\sqrt{2}}{420}
-\frac{13\sqrt{3}}{630}-\frac{13\sqrt{6}}{1260}\right)\mu_{\Sigma^{-}}\nonumber\\
& &+\left(\frac{8033}{11340}+\frac{73\sqrt{2}}{840}
-\frac{73\sqrt{3}}{1260}-\frac{73\sqrt{6}}{2520}\right)\mu_{\Xi^{0}}
+\left(\frac{1}{5}+\frac{\sqrt{2}}{10}
-\frac{\sqrt{3}}{15}-\frac{\sqrt{6}}{30}\right)\mu_{\Delta^{++}}\nonumber\\
\frac{1}{\sqrt{5}}\mu_{\Xi^{-}\Xi^{*-}}&=&\frac{49}{135}\mu_{p}+\frac{125}{324}\mu_{n}
-\frac{181}{1620}\mu_{\Sigma^{+}}
+\frac{407}{1620}\mu_{\Sigma^{-}}-\frac{109}{810}\mu_{\Xi^{0}}\nonumber\\
\frac{1}{\sqrt{15}}\mu_{\Lambda\Sigma^{*0}}&=&\left(\frac{215}{756}+\frac{17\sqrt{2}}{280}
-\frac{17\sqrt{3}}{420}-\frac{17\sqrt{6}}{840}\right)\mu_{p}
+\left(\frac{1399}{4536}+\frac{\sqrt{2}}{336}
-\frac{\sqrt{3}}{504}-\frac{\sqrt{6}}{1008}\right)\mu_{n}\nonumber\\
& &+\left(-\frac{3751}{11340}-\frac{\sqrt{2}}{840}
+\frac{\sqrt{3}}{1260}+\frac{\sqrt{6}}{2520}\right)\mu_{\Sigma^{+}}
+\left(-\frac{1093}{11340}
-\frac{13\sqrt{2}}{840}+\frac{13\sqrt{3}}{1260}+\frac{13\sqrt{6}}{2520}\right)\mu_{\Sigma^{-}}\nonumber\\
&&+\left(-\frac{5779}{22680}-\frac{73\sqrt{2}}{1680}
+\frac{73\sqrt{3}}{2520}+\frac{73\sqrt{6}}{5040}\right)\mu_{\Xi^{0}}
+\left(-\frac{1}{10}-\frac{\sqrt{2}}{20}
+\frac{\sqrt{3}}{30}+\frac{\sqrt{6}}{60}\right)\mu_{\Delta^{++}}.
\label{decsum}
\end{eqnarray}
\end{widetext}
Here we have used the additional magnetic moment
$\mu_{\Delta^{++}}$ since we have six inertia parameters ${\cal
M}$, ${\cal N}$, ${\cal N}^{\prime}$, ${\cal P}$, ${\cal Q}$ and
${\cal R}$.  We list the predictions for the magnetic moments of
the decuplet and octet baryons in Table 1, and those for the
decuplet-to-octet transition magnetic moments in Table 2, by using
their sum rules (\ref{decsum}). Here note that our prediction of
$\mu_{\Omega^{-}}$ is comparable to its experimental value
$\mu_{\Omega^{-}}^{exp}=-1.94$.  The predicted values of the
decuplet-to-octet transition magnetic moments in Table 2 are also
comparable to the previous ones obtained by using the chiral quark
soliton model~\cite{kim05}.

\begin{table}[h]
\caption{The magnetic moments, their flavor components and strange
form factors of the decuplet and octet baryons. The quantities
used as input parameters are indicated by $*$.}
\begin{center}
\begin{tabular}{lrrrrr}
\hline $B$ & $\mu_{B}^{~}$ & $\mu_{B}^{(s)}$ & $\mu_{B}^{(u)}$ &
$\mu_{B}^{(d)}$ & $F_{2B}^{(s)}$ \\ \hline
$\Delta^{++}$ & $4.52^{*}$ & $0.07$ & $4.10$ & $0.35$ & $-0.21$ \\
$\Delta^{+}$ & $2.12$ & $0.07$ & $2.50$ & $-0.45$ & $-0.21$ \\
$\Delta^{0}$ & $-0.29$ & $0.07$ & $0.89$ & $-1.25$ & $-0.21$ \\
$\Delta^{-}$ & $-2.69$ & $0.07$ & $-0.71$ & $-2.05$ & $-0.21$ \\
$\Sigma^{*+}$ & $2.63$ & $-0.49$ & $2.84$ & $0.28$ & $0.47$ \\
$\Sigma^{*0}$ & $0.08$ & $-0.49$ & $1.14$ & $-0.57$ & $0.47$ \\
$\Sigma^{*-}$ & $-2.48$ & $-0.49$ & $-0.57$ & $-1.42$ & $0.47$ \\
$\Xi^{*0}$ & $0.44$ & $-1.15$ & $1.38$ & $0.21$ & $1.45$ \\
$\Xi^{*-}$ & $-2.27$ & $-1.15$ & $-0.43$ & $-0.69$ & $1.45$ \\
$\Omega^{-}$ & $-2.06$ & $-1.92$ & $-0.29$ & $0.15$ & $2.76$ \\
$p$ & $2.79^{*}$ & $-0.25$ & $2.94$ & $0.10$ & $0.75$ \\
$n$ & $-1.91^{*}$ & $-0.25$ & $-0.19$ & $-1.47$ & $0.75$ \\
$\Sigma^{+}$ & $2.46^{*}$ & $-0.11$ & $2.72$ & $-0.15$ & $-0.67$ \\
$\Sigma^{0}$ & $0.65$ & $-0.11$ & $1.52$ & $-0.76$ & $-0.67$ \\
$\Sigma^{-}$ & $-1.16^{*}$ & $-0.11$ & $0.31$ & $-1.36$ & $-0.67$ \\
$\Xi^{0}$ & $-1.25^{*}$ & $-1.32$ & $0.25$ & $-0.18$ & $1.95$ \\
$\Xi^{-}$ & $-1.07$ & $-1.32$ & $0.37$ & $-0.12$ & $1.95$ \\
$\Lambda$ & $-0.51$ & $-0.88$ & $0.75$ & $-0.38$ & $1.64$ \\
\hline
\end{tabular}
\end{center}
\end{table}

\begin{table}[h]
\caption{The transition magnetic moments and their flavor
components for ${\bf 10}~ (J_{3}=1/2) \rightarrow {\bf
8}~(J_{3}=1/2)+\gamma$.}
\begin{center}
\begin{tabular}{lrrrr}
\hline $B_{8}B_{10}$ & $\mu_{B_{8}B_{10}}^{~}$ &
$\mu_{B_{8}B_{10}}^{(s)}$ & $\mu_{B_{8}B_{10}}^{(u)}$ &
$\mu_{B_{8}B_{10}}^{(d)}$ \\ \hline
$p\Delta^{+}$ & $-2.76$ & $-0.54$ & $-0.76$ & $-1.46$\\
$n\Delta^{0}$ & $-2.76$ & $-0.54$ & $-0.76$ & $-1.46$\\
$\Sigma^{+}\Sigma^{*+}$ & $2.24$ & $0.13$ & $2.58$ & $-0.47$\\
$\Sigma^{0}\Sigma^{*0}$ & $1.01$ & $0.13$ & $1.76$ & $-0.88$\\
$\Sigma^{-}\Sigma^{*-}$ & $-0.22$ & $0.13$ & $0.94$ & $-1.29$\\
$\Xi^{0}\Xi^{*0}$ & $2.46$ & $0.19$ & $2.72$ & $-0.45$\\
$\Xi^{-}\Xi^{*-}$ & $-0.27$ & $0.19$ & $0.90$ & $-1.36$\\
$\Lambda\Sigma^{*0}$ & $-2.46$ & $-0.54$ & $-0.56$ & $-1.36$\\
\hline
\end{tabular}
\end{center}
\end{table}

\section{Strange flavor magnetic moments of baryon decuplet}
\setcounter{equation}{0}
\renewcommand{\theequation}{\arabic{section}.\arabic{equation}}

In the SU(3) flavor symmetry broken case we decompose the EM
currents into three pieces $J^{\mu}=J^{\mu (u)}+J^{\mu (d)}+J^{\mu
(s)}$ where the q-flavor currents $J^{\mu (q)}=J^{\mu
(q)}_{CS}+J^{\mu (q)}_{FSB}$ are given by substituting the charge
operator $\hat{Q}$ with the q-flavor charge operator $\hat{Q}_{q}$
\begin{eqnarray}
J^{\mu (q)}_{CS}&=&\bar{\psi}\gamma^{\mu}\hat{Q}_{q}\psi\Theta_{B} +\left(-\frac{i%
}{2}f_{\pi}^{2}{\rm tr}(\hat{Q}_{q}l^{\mu})\right.\nn\\
&&\left.+\frac{i}{8e^{2}} {\rm tr}[\hat{Q}%
_{q},l^{\nu}][l^{\mu},l^{\nu}]+U\leftrightarrow U^{\dagger}\right)
\bar{\Theta}_{B}
\nonumber \\
& &+\frac{N_{c}}{48\pi^{2}}\epsilon^{\mu\nu\alpha\beta}{\rm tr}(\hat{Q}_{q}
l_{\nu}l_{\alpha}l_{\beta}-U\leftrightarrow U^{\dagger}) \bar{\Theta}_{B}
\nonumber \\
J^{\mu (q)}_{FSB}&=&-\frac{i}{12}f_{\pi}^{2}(\chi^{2}-1){\rm tr} ((1-\sqrt{3}%
\lambda_{8})\nn\\
&&\cdot(U\hat{Q}_{q}l^{\mu}+l^{\mu}\hat{Q}_{q}U^{\dagger})
+U\leftrightarrow U^{\dagger})\bar{\Theta}_{B}
\end{eqnarray}
to obtain the baryon magnetic moments and transition magnetic
moments in the s-flavor channel
\begin{eqnarray}
\mu_{\Delta}^{(s)}&=&-\frac{7}{48}{\cal M}+\frac{1}{12}({\cal N} -\frac{1}{2%
\sqrt{3}}{\cal N}^{\prime})\nn\\
&&+\frac{2}{21}{\cal P}+\frac{5}{168}{\cal Q}+\frac{1}{84}{\cal R}
\nonumber \\
\mu_{\Sigma^{*}}^{(s)}&=&-\frac{1}{6}{\cal M}+\frac{1}{126}{\cal P} -\frac{1%
}{126}{\cal Q}+\frac{1}{126}{\cal R}  \nonumber \\
\mu_{\Xi^{*}}^{(s)}&=&-\frac{3}{16}{\cal M}-\frac{1}{12}({\cal N} -\frac{1}{2%
\sqrt{3}}{\cal N}^{\prime})\nn\\
&&-\frac{2}{21}{\cal P}-\frac{5}{168}{\cal Q}-\frac{1}{84}{\cal R}
\nonumber \\
\mu_{\Omega}^{(s)}&=&-\frac{5}{24}{\cal M}-\frac{1}{6}({\cal N} -\frac{1}{2%
\sqrt{3}}{\cal N}^{\prime})\nn\\
&&-\frac{3}{14}{\cal P}-\frac{1}{28}{\cal Q}-\frac{1}{21}{\cal R}\nn\\
\mu_{N}^{(s)}&=&-\frac{7}{60}{\cal M}+\frac{1}{45}({\cal
N}+\frac{1}{2} {\cal N}^{\prime})\nn\\
& &+\frac{1}{45}{\cal P}+\frac{1}{90}{\cal Q}\nonumber \\
\mu_{\Sigma}^{(s)}&=&-\frac{11}{60}{\cal M}+\frac{1}{15}({\cal
N}+\frac{1}{2} {\cal N}^{\prime})\nn\\
& &+\frac{11}{135}{\cal P}+\frac{1}{54}{\cal Q}
+\frac{2}{135}{\cal R}\nonumber \\
\mu_{\Xi}^{(s)}&=&-\frac{1}{5}{\cal M}-\frac{4}{45}({\cal
N}+\frac{1}{2} {\cal N}^{\prime})\nn\\
& &-\frac{1}{9}{\cal P}-\frac{1}{45}{\cal Q}-\frac{1}{45}{\cal R}
\nonumber \\
\mu_{\Lambda}^{(s)}&=&-\frac{3}{20}{\cal M}-\frac{1}{15}({\cal
N}+\frac{1}{2} {\cal N}^{\prime})\nn\\
& &-\frac{1}{15}{\cal P}-\frac{1}{30}{\cal Q}\nn\\
\frac{1}{\sqrt{5}}\mu_{N\Delta}^{(s)}&=&-\frac{1}{6\sqrt{5}}{\cal M}\nn\\
\frac{1}{\sqrt{5}}\mu_{\Sigma\Sigma^{*}}^{(s)}&=&-\frac{1}{6\sqrt{5}}{\cal
M}+\frac{2}{45}({\cal N}+\frac{2-\sqrt{2}}{8}{\cal
N}^{\prime})\nn\\
& &+\frac{7}{135}{\cal P}+\frac{2}{135}{\cal Q}+\frac{1}{135}{\cal R}\nn\\
\frac{1}{\sqrt{5}}\mu_{\Xi\Xi^{*}}^{(s)}&=&-\frac{1}{6\sqrt{5}}{\cal
M}+\frac{2}{45}({\cal N}+\frac{2-\sqrt{2}}{8}{\cal
N}^{\prime})\nn\\
& &+\frac{1}{18}{\cal P}+\frac{1}{90}{\cal Q}+\frac{1}{90}{\cal R}\nn\\
\frac{1}{\sqrt{15}}\mu_{\Lambda\Sigma^{*0}}^{(s)}&=&
-\frac{1}{6\sqrt{15}}{\cal M}.
\end{eqnarray}

Similarly, in the u- and d-flavor channels of the adjoint
representation we obtain the baryon magnetic moments and
transition magnetic moments (\ref{mu10u}) in Appendix A.  Here one
notes that in general all the baryon decuplet and octet magnetic
moments fulfill the model independent relations in the u- and
d-flavor components and the I-spin symmetry of the isomultiplets
with the same strangeness in the s-flavor channel
\begin{equation}
\mu_{B}^{(d)}=\frac{Q_{d}}{Q_{u}}\mu_{\bar{B}}^{(u)},~~~ \mu_{B}^{(s)}=\mu_{%
\bar{B}}^{(s)}
\end{equation}
with $\bar{B}$ being the isospin conjugate baryon in the isomultiplets of
the baryon.

As in the previous sum rules of the magnetic moments
(\ref{octetsum}) and (\ref{decsum}), we find the sum rules of the
strange components of the magnetic moments and transition magnetic
moments in terms of the six experimentally known magntic moments
\begin{widetext}\begin{eqnarray}
\mu_{\Delta}^{(s)}&=&-\frac{13}{84}\mu_{p}-\frac{115}{504}\mu_{n}-\frac{83}{126}
\mu_{\Sigma^{+}}+\frac{47}{252}\mu_{\Sigma^{-}}
-\frac{463}{504}\mu_{\Xi^{0}}
+\frac{1}{6}\mu_{\Delta^{++}}
\nonumber \\
\mu_{\Sigma^{*}}^{(s)}&=&-\frac{29}{42}\mu_{p}-\frac{50}{63}\mu_{n}-\frac{23}{126}
\mu_{\Sigma^{+}}+\frac{8}{63}\mu_{\Sigma^{-}}-\frac{26}{63}\mu_{\Xi^{0}}
\nonumber \\
\mu_{\Xi^{*}}^{(s)}&=&-\frac{43}{84}\mu_{p}-\frac{445}{504}\mu_{n}-\frac{113}{126}
\mu_{\Sigma^{+}}-\frac{103}{252}\mu_{\Sigma^{-}}-\frac{433}{504}\mu_{\Xi^{0}}-\frac{1}{6}\mu_{\Delta^{++}}
\nonumber \\
\mu_{\Omega}^{(s)}&=&\frac{8}{21}\mu_{p}-\frac{125}{252}\mu_{n}-\frac{353}{126}
\mu_{\Sigma^{+}}-\frac{179}{126}\mu_{\Sigma^{-}}-\frac{569}{252}\mu_{\Xi^{0}}
-\frac{1}{3}\mu_{\Delta^{++}}\nn\\
\mu_{N}^{(s)}&=&-\frac{2}{9}\mu_{n}-\frac{7}{9}\mu_{\Sigma^{+}}-\frac{1}{9}\mu_{\Sigma^{-}}
-\frac{8}{9}\mu_{\Xi^{0}}\nonumber \\
\mu_{\Sigma}^{(s)}&=&-\frac{1}{3}\mu_{p}-\frac{5}{9}\mu_{n}-\frac{4}{9}\mu_{\Sigma^{+}}
+\frac{2}{9}\mu_{\Sigma^{-}}-\frac{8}{9}\mu_{\Xi^{0}}\nonumber \\
\mu_{\Xi}^{(s)}&=&-\frac{1}{3}\mu_{p}-\frac{2}{3}\mu_{n}-\frac{5}{3}\mu_{\Sigma^{+}}
-\frac{2}{3}\mu_{\Sigma^{-}}-\frac{4}{3}\mu_{\Xi^{0}}\nonumber \\
\mu_{\Lambda}^{(s)}&=&-\mu_{p}-\mu_{n}\nn\\
\frac{1}{\sqrt{5}}\mu_{N\Delta}^{(s)}&=&-\frac{\sqrt{5}}{15}\mu_{p}-\frac{\sqrt{5}}{9}\mu_{n}
-\frac{7\sqrt{5}}{45}\mu_{\Sigma^{+}}
-\frac{\sqrt{5}}{45}\mu_{\Sigma^{-}}-\frac{8\sqrt{5}}{45}\mu_{\Xi^{0}}\nonumber\\
\frac{1}{\sqrt{5}}\mu_{\Sigma\Sigma^{*}}^{(s)}&=&
\left(\frac{11}{210}-\frac{17\sqrt{2}}{420}
+\frac{17\sqrt{3}}{630}-\frac{\sqrt{5}}{15}
+\frac{17\sqrt{6}}{1260}\right)\mu_{p}+\left(\frac{3}{28}-\frac{\sqrt{2}}{504}
+\frac{\sqrt{3}}{756}-\frac{\sqrt{5}}{9}+\frac{\sqrt{6}}{1512}\right)\mu_{n}\nonumber\\
& &+\left(\frac{19}{210}+\frac{\sqrt{2}}{1260}
-\frac{\sqrt{3}}{1890}-\frac{7\sqrt{5}}{45}-\frac{\sqrt{6}}{3780}\right)\mu_{\Sigma^{+}}
+\left(\frac{37}{210}
+\frac{13\sqrt{2}}{1260}-\frac{13\sqrt{3}}{1890}-\frac{\sqrt{5}}{45}
-\frac{13\sqrt{6}}{3780}\right)\mu_{\Sigma^{-}}\nonumber\\
&&+\left(-\frac{13}{420}+\frac{73\sqrt{2}}{2520}
-\frac{73\sqrt{3}}{3780}-\frac{8\sqrt{5}}{45}-\frac{73\sqrt{6}}{7560}\right)\mu_{\Xi^{0}}
+\left(\frac{1}{15}+\frac{\sqrt{2}}{30}
-\frac{\sqrt{3}}{45}-\frac{\sqrt{6}}{90}\right)\mu_{\Delta^{++}}\nonumber\\
\frac{1}{\sqrt{5}}\mu_{\Xi\Xi^{*}}^{(s)}&=&
\left(-\frac{4}{35}-\frac{17\sqrt{2}}{420}
+\frac{17\sqrt{3}}{630}-\frac{\sqrt{5}}{15}
+\frac{17\sqrt{6}}{1260}\right)\mu_{p}
+\left(-\frac{1}{252}-\frac{\sqrt{2}}{504}
+\frac{\sqrt{3}}{756}-\frac{\sqrt{5}}{9}+\frac{\sqrt{6}}{1512}\right)\mu_{n}\nonumber\\
& &+\left(\frac{116}{315}+\frac{\sqrt{2}}{1260}
-\frac{\sqrt{3}}{1890}-\frac{7\sqrt{5}}{45}-\frac{\sqrt{6}}{3780}\right)\mu_{\Sigma^{+}}
+\left(\frac{181}{630}
+\frac{13\sqrt{2}}{1260}-\frac{13\sqrt{3}}{1890}-\frac{\sqrt{5}}{45}
-\frac{13\sqrt{6}}{3780}\right)\mu_{\Sigma^{-}}\nonumber\\
&&+\left(\frac{241}{1260}+\frac{73\sqrt{2}}{2520}
-\frac{73\sqrt{3}}{3780}-\frac{8\sqrt{5}}{45}-\frac{73\sqrt{6}}{7560}\right)\mu_{\Xi^{0}}
+\left(\frac{1}{15}+\frac{\sqrt{2}}{30}
-\frac{\sqrt{3}}{45}-\frac{\sqrt{6}}{90}\right)\mu_{\Delta^{++}}\nonumber\\
\frac{1}{\sqrt{15}}\mu_{\Lambda\Sigma^{*0}}^{(s)}&=&
-\frac{\sqrt{15}}{45}\mu_{p}-\frac{\sqrt{15}}{27}\mu_{n}
-\frac{7\sqrt{15}}{135}\mu_{\Sigma^{+}}
-\frac{\sqrt{15}}{135}\mu_{\Sigma^{-}}-\frac{8\sqrt{15}}{135}\mu_{\Xi^{0}}.
\label{mu10ssum}
\end{eqnarray}
\end{widetext}
Similarly, the sum rules of the u- and d-flavor components of the
magnetic moments and transition magnetic moments are given by
(\ref{mu10usum}) in Appendix A. Here one notes that the flavor
components of the transition magnetic moments $\mu_{p\Delta^{+}}$
and $\mu_{n\Delta^{0}}$ satisfy the identities \beq
\mu_{p\Delta^{+}}^{(q)}=\mu_{n\Delta^{0}}^{(q)},~~~(q=u, d,
s),\label{uds} \eeq to yield $\mu_{p\Delta^{+}}=\mu_{n\Delta^{0}}$
as in (\ref{decsum}).  The identities (\ref{uds}) are consistent
with the previous ones in Refs.~\cite{jenkins94plb,lebed04}. We
list the predictions for the u-, d- and s-flavor components of the
decuplet and octet magnetic moments in Table 1, and those for the
decuplet-to-octet transition magnetic moments in Table 2, by using
the sum rules (\ref{mu10ssum}) and (\ref{mu10usum}).

Next, the form factors of the decuplet baryons, with internal
structure, are defined by the matrix elements of the EM currents
\bea
\langle p+q|J^{\mu}|p\rangle&=&\bar{u}(p+q)(\gamma^{\mu}F_{1B}(q^{2})\nn\\
&&~~+\frac{i%
}{2m_{B}}\sigma^{\mu\nu}q^{\nu}F_{2B}(q^{2}))u(p) \eea where
$u(p)$ is the spinor of the baryons and $q$ is the momentum
transfer. Using the s-flavor charge operator in the EM currents as
before, in the limit of zero momentum transfer, one can obtain the
strange form factors of baryon decuplet and octet
\begin{eqnarray}
F_{1B}^{(s)}(0)&=&S  \nonumber \\
F_{2B}^{(s)}(0)&=&-3\mu_{B}^{(s)}-S \label{sforms}
\end{eqnarray}
in terms of the strangeness quantum number of the baryon $S(=1-Y)$
($Y$: hypercharge) and the strange components of the baryon
decuplet and octet magnetic moments $\mu_{B}^{(s)}$. The
predictions for the strange form factors of the decuplet and octet
baryons are listed in Table 1 by using the relation
(\ref{sforms}).

\section{Conclusions}
\setcounter{equation}{0}
\renewcommand{\theequation}{\arabic{section}.\arabic{equation}}

In summary, we have derived sum rules for the baryon decuplet and
octet magnetic moments and the decuplet-to-octet transition
magnetic moments, in the chiral models with the SU(3) flavor
group. These sum rules are explicitly constructed in terms of the
six experimentally known baryon magnetic moments $\mu_{p}$,
$\mu_{n}$, $\mu_{\Sigma^{+}}$, $\mu_{\Sigma^{-}}$, $\mu_{\Xi^{0}}$
and $\mu_{\Delta^{++}}$ to yield the theoretical predictions for
the remnant baryon magnetic moments. Especially in case of using
the experimental data for the six baryon magnetic moments as input
data of the sum rules, the predicted value of $\mu_{\Omega^{-}}$
is comparable to its experimental datum.

Next, we have extended the above algorithm to flavor components
and strange form factors of the baryon decuplet and octet magnetic
moments to find their sum rules in terms of the six baryon
magnetic moments.  The sum rules for the decuplet-to-octet
transition magnetic moments and their flavor components have been
also obtained. It is also shown that all the baryon decuplet and
octet magnetic moments fulfill the model independent relations in
the u- and d-flavor components and the I-spin symmetry of the
isomultiplets with the same strangeness in the s-flavor channel.
Moreover, the s-flavor components of the decuplet-to-octet
transition magnetic moments respect the I-spin symmetry of the
isomultiplets with the same strangeness. However, some transition
magnetic moments such as $\mu_{p\Delta^{+}}^{(u)}$,
$\mu_{n\Delta^{0}}^{(u)}$ and $\mu_{\Lambda\Sigma^{*0}}^{(u)}$ do
not satisfy the model independent relations in the u- and d-flavor
components.  It is also interesting to see that
$\mu_{p\Delta^{+}}^{(q)}$ and $\mu_{n\Delta^{0}}^{(q)}$ are equal
to each other in the q-flavor $(q=u,d,s)$ channels, which are
consistent with the previous results~\cite{jenkins94plb,lebed04}.

It would be desirable if the SU(3) representation mixing effects
on the baryon magnetic moments~\cite{hongpr01} can be investigated
by exploiting the multiquark structure associated with the highly
nontrivial nonlinear symmetry breaking terms.

\acknowledgments The author would like to deeply thank R.D.
McKeown for the warm hospitality at Kellogg Radiation Laboratory,
Caltech where a part of this work has been done. He is also
grateful to G.E. Brown, H.C. Kim, R.D. McKeown, D.P. Min, B.Y.
Park, M. Ramsey-Musolf, M. Rho and P. Wang for helpful discussions
and encouragements. This work was supported by the Korea Research
Foundation Grant funded by the Korean Government (MOEHRD, Basic
Research Promotion Fund), Grant No. KRF-2006-331-C00071, and by
the Korea Research Council of Fundamental Science and Technology
(KRCF), Grant No. C-RESEARCH-2006-11-NIMS.

\appendix
\section{}
\setcounter{equation}{0}
\renewcommand{\theequation}{A.\arabic{equation}}

In the u- and d-flavor channels of the adjoint representation, the
baryon magnetic moments and transition magnetic moments are given
by \bea
\mu_{\Delta^{++}}^{(u)}&=&-2\mu_{\Delta^{-}}^{(d)}=\frac{5}{12}{\cal
M}+\frac{1}{3}({\cal N}-\frac{1} {2\sqrt{3}}{\cal
N}^{\prime})\nn\\
&&+\frac{2}{7}{\cal P}-\frac{1}{28}{\cal Q}-\frac{1}{21}{\cal R}
\nonumber \\
\mu_{\Delta^{+}}^{(u)}&=&-2\mu_{\Delta^{0}}^{(d)}=\frac{3}{8}{\cal
M}+\frac{1}{6}({\cal N}-\frac{1} {2\sqrt{3}}{\cal
N}^{\prime})\nn\\
&&+\frac{10}{63}{\cal P}+\frac{1}{126}{\cal Q}-\frac{1}{126}{\cal
R}
\nonumber \\
\mu_{\Delta^{0}}^{(u)}&=&-2\mu_{\Delta^{+}}^{(d)}=\frac{1}{3}{\cal
M}+\frac{2}{63}{\cal P}
+\frac{13}{252}{\cal Q}+\frac{2}{63}{\cal R}  \nonumber \\
\mu_{\Delta^{-}}^{(u)}&=&-2\mu_{\Delta^{++}}^{(d)}=\frac{7}{24}{\cal
M}-\frac{1}{6}({\cal N}-\frac{1} {2\sqrt{3}}{\cal
N}^{\prime})\nn\\
&&-\frac{2}{21}{\cal P}+\frac{2}{21}{\cal Q}+\frac{1}{14}{\cal R}
\nonumber \\
\mu_{\Sigma^{*+}}^{(u)}&=&-2\mu_{\Sigma^{*-}}^{(d)}=\frac{3}{8}{\cal
M}+\frac{1}{6}({\cal N}-\frac{1} {2\sqrt{3}}{\cal
N}^{\prime})\nn\\
&&+\frac{19}{126}{\cal P}-\frac{17}{252}{\cal Q}
-\frac{1}{63}{\cal R}\nonumber \\
\mu_{\Sigma^{*0}}^{(u)}&=&-2\mu_{\Sigma^{*0}}^{(d)}=\frac{1}{3}{\cal
M}+\frac{1}{126}{\cal P}
-\frac{1}{126}{\cal Q}+\frac{1}{126}{\cal R}  \nonumber \\
\mu_{\Sigma^{*-}}^{(u)}&=&-2\mu_{\Sigma^{*+}}^{(d)}=\frac{7}{24}{\cal
M}-\frac{1}{6}({\cal N}-\frac{1} {2\sqrt{3}}{\cal
N}^{\prime})\nn\\
&&-\frac{17}{126}{\cal P}+\frac{13}{252}{\cal Q}
+\frac{2}{63}{\cal R}\nonumber \\
\mu_{\Xi^{*0}}^{(u)}&=&-2\mu_{\Xi^{*-}}^{(d)}=\frac{1}{3}{\cal
M}-\frac{1}{63}{\cal P}
-\frac{17}{252}{\cal Q}-\frac{1}{63}{\cal R}  \nonumber \\
\mu_{\Xi^{*-}}^{(u)}&=&-2\mu_{\Xi^{*0}}^{(d)}=\frac{7}{24}{\cal
M}-\frac{1}{6}({\cal N}-\frac{1} {2\sqrt{3}}{\cal
N}^{\prime})\nn\\
&&-\frac{11}{63}{\cal P}+\frac{1}{126}{\cal Q}
-\frac{1}{126}{\cal R}\nonumber\\
\mu_{\Omega^{-}}^{(u)}&=&-2\mu_{\Omega^{-}}^{(d)}=\frac{7}{24}{\cal
M}-\frac{1}{6}({\cal N}-\frac{1} {2\sqrt{3}}{\cal
N}^{\prime})\nn\\
&&-\frac{3}{14}{\cal P}-\frac{1}{28}{\cal Q}-\frac{1}{21}{\cal
R}\nn\\
\mu_{p}^{(u)}&=&-2\mu_{n}^{(d)}=\frac{2}{5}{\cal
M}+\frac{8}{45}({\cal N}+\frac{1}{2}{\cal
N}^{\prime})\nn\\
&&+\frac{16}{135}{\cal P}-\frac{4}{135}{\cal Q}-\frac{2}{135}{\cal
R}
\nonumber \\
\mu_{n}^{(u)}&=&-2\mu_{p}^{(d)}=\frac{11}{30}{\cal
M}-\frac{2}{15}({\cal N}+\frac{1}{2}{\cal
N}^{\prime})\nn\\
&&-\frac{2}{27}{\cal P}+\frac{7}{135}{\cal Q}+\frac{2}{135}{\cal
R}
\nonumber \\
\mu_{\Sigma^{+}}^{(u)}&=&-2\mu_{\Sigma^{-}}^{(d)}=\frac{2}{5}{\cal
M}+\frac{8}{45}({\cal N}+\frac{1}{2}{\cal
N}^{\prime})\nn\\
&&+\frac{26}{135}{\cal P}-\frac{2}{135}{\cal Q}
-\frac{4}{135}{\cal R}\nonumber \\
\mu_{\Sigma^{0}}^{(u)}&=&-2\mu_{\Sigma^{0}}^{(d)}=\frac{19}{60}{\cal
M}+\frac{1}{15}({\cal N}+\frac{1}{2}{\cal
N}^{\prime})\nn\\
& &+\frac{11}{135}{\cal P}
+\frac{1}{54}{\cal Q}+\frac{2}{135}{\cal R}  \nonumber \\
\mu_{\Sigma^{-}}^{(u)}&=&-2\mu_{\Sigma^{+}}^{(d)}=\frac{7}{30}{\cal
M}-\frac{2}{45}({\cal N}+\frac{1}{2}{\cal
N}^{\prime})\nn\\
&&-\frac{4}{135}{\cal P}+\frac{7}{135}{\cal Q}
+\frac{8}{135}{\cal R}\nonumber \\
\mu_{\Xi^{0}}^{(u)}&=&-2\mu_{\Xi^{-}}^{(d)}=\frac{11}{30}{\cal
M}-\frac{2}{15}({\cal N}+\frac{1}{2}{\cal
N}^{\prime})\nn\\
& &-\frac{22}{135}{\cal P}
-\frac{2}{135}{\cal Q}+\frac{2}{135}{\cal R}  \nonumber \\
\mu_{\Xi^{-}}^{(u)}&=&-2\mu_{\Xi^{0}}^{(d)}=\frac{7}{30}{\cal
M}-\frac{2}{45}({\cal N}+\frac{1}{2}{\cal
N}^{\prime})\nn\\
&&-\frac{8}{135}{\cal P}-\frac{4}{135}{\cal Q}
-\frac{8}{135}{\cal R}\nonumber\\
\mu_{\Lambda}^{(u)}&=&-2\mu_{\Lambda}^{(d)}=\frac{7}{20}{\cal
M}-\frac{1}{15}({\cal N}+\frac{1}{2} {\cal
N}^{\prime})\nn\\
& &-\frac{1}{15}{\cal
P}-\frac{1}{30}{\cal Q}\nn\\
\frac{1}{\sqrt{5}}\mu_{p\Delta^{+}}^{(u)}&=&\frac{1}{\sqrt{5}}\mu_{n\Delta^{0}}^{(u)}\nn\\
&=&\frac{1}{3\sqrt{5}}{\cal M}-\frac{4}{45}({\cal
N}+\frac{2-\sqrt{2}}{8}{\cal
N}^{\prime})\nn\\
& &-\frac{8}{135}{\cal P}+\frac{7}{270}{\cal Q}+\frac{79}{4050}{\cal R}\nn\\
\frac{1}{\sqrt{5}}\mu_{\Sigma^{+}\Sigma^{*+}}^{(u)}&=&-\frac{2}{\sqrt{5}}\mu_{\Sigma^{-}\Sigma^{*-}}^{(d)}\nn\\
&=&\frac{1}{3\sqrt{5}}{\cal M}+\frac{4}{45}({\cal
N}+\frac{2-\sqrt{2}}{8}{\cal
N}^{\prime})\nn\\
& &+\frac{13}{135}{\cal P}+\frac{1}{270}{\cal
Q}-\frac{1}{150}{\cal R}\nn\\
\frac{1}{\sqrt{5}}\mu_{\Sigma^{0}\Sigma^{*0}}^{(u)}&=&-\frac{2}{\sqrt{5}}\mu_{\Sigma^{0}\Sigma^{*0}}^{(d)}\nn\\
&=&\frac{1}{3\sqrt{5}}{\cal M}+\frac{2}{45}({\cal
N}+\frac{2-\sqrt{2}}{8}{\cal
N}^{\prime})\nn\\
& &+\frac{7}{135}{\cal P}+\frac{2}{135}{\cal
Q}+\frac{1}{135}{\cal R}\nn\\
\frac{1}{\sqrt{5}}\mu_{\Sigma^{-}\Sigma^{*-}}^{(u)}&=&-\frac{2}{\sqrt{5}}\mu_{\Sigma^{+}\Sigma^{*+}}^{(d)}\nn\\
&=&\frac{1}{3\sqrt{5}}{\cal M}+\frac{1}{135}{\cal
P}+\frac{7}{270}{\cal
Q}+\frac{29}{1350}{\cal R}\nn\\
\frac{1}{\sqrt{5}}\mu_{\Xi^{0}\Xi^{*0}}^{(u)}&=&-\frac{2}{\sqrt{5}}\mu_{\Xi^{-}\Xi^{*-}}^{(d)}\nn\\
&=&\frac{1}{3\sqrt{5}}{\cal M}+\frac{4}{45}({\cal
N}+\frac{2-\sqrt{2}}{8}{\cal N}^{\prime})\nn\\
& &+\frac{14}{135}{\cal P}-\frac{1}{270}{\cal
Q}+\frac{23}{4050}{\cal R}\nn\\
\frac{1}{\sqrt{5}}\mu_{\Xi^{-}\Xi^{*-}}^{(u)}&=&-\frac{2}{\sqrt{5}}\mu_{\Xi^{0}\Xi^{*0}}^{(d)}\nn\\
&=&\frac{1}{3\sqrt{5}}{\cal M}+\frac{1}{135}{\cal
P}+\frac{7}{270}{\cal
Q}+\frac{67}{4050}{\cal R}\nn\\
\frac{1}{\sqrt{15}}\mu_{\Lambda\Sigma^{*0}}^{(u)}&=&\frac{1}{3\sqrt{15}}{\cal
M}-\frac{2}{45}({\cal
N}+\frac{2-\sqrt{2}}{8}{\cal N}^{\prime})\nn\\
& &-\frac{1}{27}{\cal P}+\frac{2}{135}{\cal
Q}+\frac{11}{2025}{\cal R}\nn\\
\frac{1}{\sqrt{5}}\mu_{p\Delta^{+}}^{(d)}&=&\frac{1}{\sqrt{5}}\mu_{n\Delta^{0}}^{(d)}\nn\\
&=&-\frac{1}{6\sqrt{5}}{\cal M}-\frac{2}{45}({\cal
N}+\frac{2-\sqrt{2}}{8}{\cal
N}^{\prime})\nn\\
& &-\frac{4}{135}{\cal P}+\frac{7}{540}{\cal Q}+\frac{79}{8100}{\cal R}\nn\\
\frac{1}{\sqrt{15}}\mu_{\Lambda\Sigma^{*0}}^{(d)}&=&-\frac{1}{6\sqrt{15}}{\cal
M}-\frac{1}{45}({\cal
N}+\frac{2-\sqrt{2}}{8}{\cal N}^{\prime})\nn\\
& &-\frac{1}{54}{\cal P}+\frac{1}{135}{\cal
Q}+\frac{11}{4050}{\cal R}. \label{mu10u} \eea

Next, we list the sum rules of the u- and d-flavor components of
the magnetic moments and transition magnetic moments
\begin{widetext}\bea
\mu_{\Delta^{++}}^{(u)}&=&\frac{2}{3}\mu_{p}+\frac{10}{9}\mu_{n}+\frac{14}{9}
\mu_{\Sigma^{+}}+\frac{2}{9}\mu_{\Sigma^{-}}
+\frac{16}{9}\mu_{\Xi^{0}}+\frac{2}{3}\mu_{\Delta^{++}}
\nonumber \\
\mu_{\Delta^{+}}^{(u)}&=&\frac{11}{14}\mu_{p}+\frac{335}{252}\mu_{n}+\frac{103}{63}
\mu_{\Sigma^{+}}+\frac{53}{126}\mu_{\Sigma^{-}}
+\frac{443}{252}\mu_{\Xi^{0}}+\frac{1}{3}\mu_{\Delta^{++}}
\nonumber \\
\mu_{\Delta^{0}}^{(u)}&=&\frac{19}{21}\mu_{p}+\frac{65}{42}\mu_{n}+\frac{12}{7}
\mu_{\Sigma^{+}}+\frac{13}{21}\mu_{\Sigma^{-}}
+\frac{73}{42}\mu_{\Xi^{0}}\nonumber \\
\mu_{\Delta^{-}}^{(u)}&=&\frac{43}{42}\mu_{p}+\frac{445}{252}\mu_{n}+\frac{113}{63}
\mu_{\Sigma^{+}}+\frac{103}{126}\mu_{\Sigma^{-}}
+\frac{433}{252}\mu_{\Xi^{0}}-\frac{1}{3}\mu_{\Delta^{++}}
\nonumber \\
\mu_{\Sigma^{*+}}^{(u)}&=&-\frac{11}{21}\mu_{p}+\frac{5}{28}\mu_{n}+\frac{137}{42}
\mu_{\Sigma^{+}}+\frac{31}{42}\mu_{\Sigma^{-}}
+\frac{271}{84}\mu_{\Xi^{0}}+\frac{1}{3}\mu_{\Delta^{++}}
\nonumber \\
\mu_{\Sigma^{*0}}^{(u)}&=&\frac{13}{42}\mu_{p}+\frac{55}{63}\mu_{n}+\frac{271}{126}
\mu_{\Sigma^{+}}+\frac{29}{63}\mu_{\Sigma^{-}}
+\frac{142}{63}\mu_{\Xi^{0}}\nonumber \\
\mu_{\Sigma^{*-}}^{(u)}&=&\frac{8}{7}\mu_{p}+\frac{395}{252}\mu_{n}+\frac{131}{126}
\mu_{\Sigma^{+}}+\frac{23}{126}\mu_{\Sigma^{-}}
+\frac{323}{252}\mu_{\Xi^{0}}-\frac{1}{3}\mu_{\Delta^{++}}
\nonumber \\
\mu_{\Xi^{*0}}^{(u)}&=&-\frac{2}{7}\mu_{p}+\frac{25}{126}\mu_{n}+\frac{163}{63}
\mu_{\Sigma^{+}}+\frac{19}{63}\mu_{\Sigma^{-}}
+\frac{349}{126}\mu_{\Xi^{0}}
\nonumber \\
\mu_{\Xi^{*-}}^{(u)}&=&\frac{53}{42}\mu_{p}+\frac{115}{84}\mu_{n}+\frac{2}{7}
\mu_{\Sigma^{+}}-\frac{19}{42}\mu_{\Sigma^{-}}
+\frac{71}{84}\mu_{\Xi^{0}}-\frac{1}{3}\mu_{\Delta^{++}}
\nonumber\\
\mu_{\Omega^{-}}^{(u)}&=&\frac{29}{21}\mu_{p}+\frac{295}{252}\mu_{n}-\frac{59}{126}
\mu_{\Sigma^{+}}-\frac{137}{126}\mu_{\Sigma^{-}}
+\frac{103}{252}\mu_{\Xi^{0}}-\frac{1}{3}\mu_{\Delta^{++}}\nn\\
\mu_{p}^{(u)}&=&\frac{4}{3}\mu_{p}+\frac{10}{9}\mu_{n}+\frac{14}{9}
\mu_{\Sigma^{+}}+\frac{2}{9}\mu_{\Sigma^{-}}
+\frac{16}{9}\mu_{\Xi^{0}}\nonumber \\
\mu_{n}^{(u)}&=&\frac{2}{3}\mu_{p}+\frac{16}{9}\mu_{n}+\frac{14}{9}
\mu_{\Sigma^{+}}+\frac{2}{9}\mu_{\Sigma^{-}}
+\frac{16}{9}\mu_{\Xi^{0}}\nonumber \\
\mu_{\Sigma^{+}}^{(u)}&=&\frac{2}{3}\mu_{p}+\frac{10}{9}\mu_{n}+\frac{20}{9}
\mu_{\Sigma^{+}}+\frac{2}{9}\mu_{\Sigma^{-}}
+\frac{16}{9}\mu_{\Xi^{0}}\nonumber \\
\mu_{\Sigma^{0}}^{(u)}&=&\frac{2}{3}\mu_{p}+\frac{10}{9}\mu_{n}+\frac{17}{9}
\mu_{\Sigma^{+}}+\frac{5}{9}\mu_{\Sigma^{-}}
+\frac{16}{9}\mu_{\Xi^{0}}\nonumber \\
\mu_{\Sigma^{-}}^{(u)}&=&\frac{2}{3}\mu_{p}+\frac{10}{9}\mu_{n}+\frac{14}{9}
\mu_{\Sigma^{+}}+\frac{8}{9}\mu_{\Sigma^{-}}
+\frac{16}{9}\mu_{\Xi^{0}}\nonumber \\
\mu_{\Xi^{0}}^{(u)}&=&\frac{2}{3}\mu_{p}+\frac{10}{9}\mu_{n}+\frac{14}{9}
\mu_{\Sigma^{+}}+\frac{2}{9}\mu_{\Sigma^{-}}
+\frac{22}{9}\mu_{\Xi^{0}}\nonumber \\
\mu_{\Xi^{-}}^{(u)}&=&\frac{2}{3}\mu_{p}+\frac{8}{9}\mu_{n}-\frac{2}{9}
\mu_{\Sigma^{+}}-\frac{8}{9}\mu_{\Sigma^{-}}
+\frac{2}{9}\mu_{\Xi^{0}}\nonumber\\
\mu_{\Lambda}^{(u)}&=&\frac{2}{3}\mu_{n}+\frac{7}{3}
\mu_{\Sigma^{+}}+\frac{1}{3}\mu_{\Sigma^{-}}\nn
+\frac{8}{3}\mu_{\Xi^{0}}\nn\\
\frac{1}{\sqrt{5}}\mu_{p\Delta^{+}}^{(u)}&=&\frac{1}{\sqrt{5}}\mu_{n\Delta^{0}}^{(u)}\nonumber\\
&=&\left(-\frac{2}{567}+\frac{17\sqrt{2}}{210}
-\frac{17\sqrt{3}}{315}+\frac{2\sqrt{5}}{15}
-\frac{17\sqrt{6}}{630}\right)\mu_{p}+\left(\frac{346}{1701}+\frac{\sqrt{2}}{252}
-\frac{\sqrt{3}}{378}+\frac{2\sqrt{5}}{9}-\frac{\sqrt{6}}{756}\right)\mu_{n}\nonumber\\
& &+\left(\frac{3383}{17010}-\frac{\sqrt{2}}{630}
+\frac{\sqrt{3}}{945}+\frac{14\sqrt{5}}{45}+\frac{\sqrt{6}}{1890}\right)\mu_{\Sigma^{+}}+\left(\frac{2189}{17010}
-\frac{13\sqrt{2}}{630}+\frac{13\sqrt{3}}{945}+\frac{2\sqrt{5}}{45}
+\frac{13\sqrt{6}}{1890}\right)\mu_{\Sigma^{-}}\nonumber\\
&&+\left(\frac{2131}{17010}-\frac{73\sqrt{2}}{1260}
+\frac{73\sqrt{3}}{1890}+\frac{16\sqrt{5}}{45}+\frac{73\sqrt{6}}{3780}\right)\mu_{\Xi^{0}}
+\left(-\frac{2}{15}-\frac{\sqrt{2}}{15}
+\frac{2\sqrt{3}}{45}+\frac{\sqrt{6}}{45}\right)\mu_{\Delta^{++}}
\nonumber\\
\frac{1}{\sqrt{5}}\mu_{\Sigma^{+}\Sigma^{*+}}^{(u)}&=&
\left(-\frac{83}{945}-\frac{17\sqrt{2}}{210}
+\frac{17\sqrt{3}}{315}+\frac{2\sqrt{5}}{15}
+\frac{17\sqrt{6}}{630}\right)\mu_{p}+\left(-\frac{1}{567}-\frac{\sqrt{2}}{252}
+\frac{\sqrt{3}}{378}+\frac{2\sqrt{5}}{9}+\frac{\sqrt{6}}{756}\right)\mu_{n}\nonumber\\
& &+\left(\frac{131}{1134}+\frac{\sqrt{2}}{630}
-\frac{\sqrt{3}}{945}+\frac{14\sqrt{5}}{45}-\frac{\sqrt{6}}{1890}\right)\mu_{\Sigma^{+}}+\left(\frac{107}{1134}
+\frac{13\sqrt{2}}{630}-\frac{13\sqrt{3}}{945}+\frac{2\sqrt{5}}{45}
-\frac{13\sqrt{6}}{1890}\right)\mu_{\Sigma^{-}}\nonumber\\
&&+\left(-\frac{589}{5670}+\frac{73\sqrt{2}}{1260}
-\frac{73\sqrt{3}}{1890}+\frac{16\sqrt{5}}{45}-\frac{73\sqrt{6}}{3780}\right)\mu_{\Xi^{0}}
+\left(\frac{2}{15}+\frac{\sqrt{2}}{15}
-\frac{2\sqrt{3}}{45}-\frac{\sqrt{6}}{45}\right)\mu_{\Delta^{++}}
\nonumber\\
\frac{1}{\sqrt{5}}\mu_{\Sigma^{0}\Sigma^{*0}}^{(u)}&=&
\left(\frac{11}{210}-\frac{17\sqrt{2}}{420}
+\frac{17\sqrt{3}}{630}+\frac{2\sqrt{5}}{15}
+\frac{17\sqrt{6}}{1260}\right)\mu_{p}+\left(\frac{3}{28}-\frac{\sqrt{2}}{504}
+\frac{\sqrt{3}}{756}+\frac{2\sqrt{5}}{9}+\frac{\sqrt{6}}{1512}\right)\mu_{n}\nonumber\\
& &+\left(\frac{19}{210}+\frac{\sqrt{2}}{1260}
-\frac{\sqrt{3}}{1890}+\frac{14\sqrt{5}}{45}-\frac{\sqrt{6}}{3780}\right)\mu_{\Sigma^{+}}+\left(\frac{37}{210}
+\frac{13\sqrt{2}}{1260}-\frac{13\sqrt{3}}{1890}+\frac{2\sqrt{5}}{45}-\frac{13\sqrt{6}}{3780}\right)
\mu_{\Sigma^{-}}\nonumber\\
&&+\left(-\frac{13}{420}+\frac{73\sqrt{2}}{2520}
-\frac{73\sqrt{3}}{3780}+\frac{16\sqrt{5}}{45}-\frac{73\sqrt{6}}{7560}\right)\mu_{\Xi^{0}}
+\left(\frac{1}{15}+\frac{\sqrt{2}}{30}
-\frac{\sqrt{3}}{45}-\frac{\sqrt{6}}{90}\right)\mu_{\Delta^{++}}\nonumber\\
\frac{1}{\sqrt{5}}\mu_{\Sigma^{-}\Sigma^{*-}}^{(u)}&=&
\left(\frac{26}{135}+\frac{2\sqrt{5}}{15}\right)\mu_{p}+\left(\frac{35}{162}+\frac{2\sqrt{5}}{9}\right)\mu_{n}
+\left(\frac{53}{810}+\frac{14\sqrt{5}}{45}\right)\mu_{\Sigma^{+}}+\left(\frac{209}{810}
+\frac{2\sqrt{5}}{45}\right)\mu_{\Sigma^{-}}\nonumber\\
&&+\left(\frac{17}{405}+\frac{16\sqrt{5}}{45}\right)\mu_{\Xi^{0}}\nonumber\\
\frac{1}{\sqrt{5}}\mu_{\Xi^{0}\Xi^{*0}}^{(u)}&=&
\left(-\frac{1334}{2835}-\frac{17\sqrt{2}}{210}
+\frac{17\sqrt{3}}{315}+\frac{2\sqrt{5}}{15}
+\frac{17\sqrt{6}}{630}\right)\mu_{p}+\left(-\frac{451}{1701}-\frac{\sqrt{2}}{252}
+\frac{\sqrt{3}}{378}+\frac{2\sqrt{5}}{9}+\frac{\sqrt{6}}{756}\right)\mu_{n}\nonumber\\
& &+\left(\frac{2759}{3402}+\frac{\sqrt{2}}{630}
-\frac{\sqrt{3}}{945}+\frac{14\sqrt{5}}{45}-\frac{\sqrt{6}}{1890}\right)\mu_{\Sigma^{+}}
+\left(\frac{1385}{3402}
+\frac{13\sqrt{2}}{630}-\frac{13\sqrt{3}}{945}+\frac{2\sqrt{5}}{45}
-\frac{13\sqrt{6}}{1890}\right)\mu_{\Sigma^{-}}\nonumber\\
&&+\left(\frac{8033}{17010}+\frac{73\sqrt{2}}{1260}
-\frac{73\sqrt{3}}{1890}+\frac{16\sqrt{5}}{45}-\frac{73\sqrt{6}}{3780}\right)\mu_{\Xi^{0}}
+\left(\frac{2}{15}+\frac{\sqrt{2}}{15}
-\frac{2\sqrt{3}}{45}-\frac{\sqrt{6}}{45}\right)\mu_{\Delta^{++}}
\nonumber\\
\frac{1}{\sqrt{5}}\mu_{\Xi^{-}\Xi^{*-}}^{(u)}&=&
\left(\frac{98}{405}+\frac{2\sqrt{5}}{15}\right)\mu_{p}+\left(\frac{125}{486}+\frac{2\sqrt{5}}{9}\right)\mu_{n}
+\left(-\frac{181}{2430}+\frac{14\sqrt{5}}{45}\right)\mu_{\Sigma^{+}}+\left(\frac{407}{2430}
+\frac{2\sqrt{5}}{45}\right)\mu_{\Sigma^{-}}\nonumber\\
&&+\left(-\frac{109}{1215}+\frac{16\sqrt{5}}{45}\right)\mu_{\Xi^{0}}
\nonumber\\
\frac{1}{\sqrt{15}}\mu_{\Lambda\Sigma^{*0}}^{(u)}&=&
\left(\frac{215}{1134}+\frac{17\sqrt{2}}{420}
-\frac{17\sqrt{3}}{630}-\frac{17\sqrt{6}}{1260}
+\frac{2\sqrt{15}}{45}\right)\mu_{p}
+\left(\frac{1399}{6804}+\frac{\sqrt{2}}{504}
-\frac{\sqrt{3}}{756}-\frac{\sqrt{6}}{1512}+\frac{2\sqrt{15}}{27}\right)\mu_{n}\nonumber\\
& &+\left(-\frac{3751}{17010}-\frac{\sqrt{2}}{1260}
+\frac{\sqrt{3}}{1890}+\frac{\sqrt{6}}{3780}+\frac{14\sqrt{15}}{135}\right)\mu_{\Sigma^{+}}\nonumber\\
& &+\left(-\frac{1093}{17010}
-\frac{13\sqrt{2}}{1260}+\frac{13\sqrt{3}}{1890}+\frac{13\sqrt{6}}{3780}
+\frac{2\sqrt{15}}{135}\right)\mu_{\Sigma^{-}}\nonumber\\
&&+\left(-\frac{5779}{34020}-\frac{73\sqrt{2}}{2520}
+\frac{73\sqrt{3}}{3780}+\frac{73\sqrt{6}}{7560}+\frac{16\sqrt{15}}{135}\right)\mu_{\Xi^{0}}
+\left(-\frac{1}{15}-\frac{\sqrt{2}}{30}
+\frac{\sqrt{3}}{45}+\frac{\sqrt{6}}{90}\right)\mu_{\Delta^{++}}\nonumber\\
\frac{1}{\sqrt{5}}\mu_{p\Delta^{+}}^{(d)}&=&\frac{1}{\sqrt{5}}\mu_{n\Delta^{0}}^{(d)}\nonumber\\
&=&\left(-\frac{1}{567}+\frac{17\sqrt{2}}{420}
-\frac{17\sqrt{3}}{630}-\frac{\sqrt{5}}{15}
-\frac{17\sqrt{6}}{1260}\right)\mu_{p}
+\left(\frac{173}{1701}+\frac{\sqrt{2}}{504}
-\frac{\sqrt{3}}{756}-\frac{\sqrt{5}}{9}-\frac{\sqrt{6}}{1512}\right)\mu_{n}\nonumber\\
& &+\left(\frac{3383}{34020}-\frac{\sqrt{2}}{1260}
+\frac{\sqrt{3}}{1890}-\frac{7\sqrt{5}}{45}+\frac{\sqrt{6}}{3780}\right)\mu_{\Sigma^{+}}
+\left(\frac{2189}{34020}
-\frac{13\sqrt{2}}{1260}+\frac{13\sqrt{3}}{1890}-\frac{\sqrt{5}}{45}
+\frac{13\sqrt{6}}{3780}\right)\mu_{\Sigma^{-}}\nonumber\\
&&+\left(\frac{2131}{34020}-\frac{73\sqrt{2}}{2520}
+\frac{73\sqrt{3}}{3780}-\frac{8\sqrt{5}}{45}+\frac{73\sqrt{6}}{7560}\right)\mu_{\Xi^{0}}
+\left(-\frac{1}{15}-\frac{\sqrt{2}}{30}
+\frac{\sqrt{3}}{45}+\frac{\sqrt{6}}{90}\right)\mu_{\Delta^{++}}
\nonumber\\
\frac{1}{\sqrt{15}}\mu_{\Lambda\Sigma^{*0}}^{(d)}&=&
\left(\frac{215}{2268}+\frac{17\sqrt{2}}{840}
-\frac{17\sqrt{3}}{1260}-\frac{17\sqrt{6}}{2520}
-\frac{\sqrt{15}}{45}\right)\mu_{p}
+\left(\frac{1399}{13608}+\frac{\sqrt{2}}{1008}
-\frac{\sqrt{3}}{1512}-\frac{\sqrt{6}}{3024}-\frac{\sqrt{15}}{27}\right)\mu_{n}\nonumber\\
& &+\left(-\frac{3751}{34020}-\frac{\sqrt{2}}{2520}
+\frac{\sqrt{3}}{3780}+\frac{\sqrt{6}}{7560}-\frac{7\sqrt{15}}{135}\right)\mu_{\Sigma^{+}}\nonumber\\
& &+\left(-\frac{1093}{34020}
-\frac{13\sqrt{2}}{2520}+\frac{13\sqrt{3}}{3780}+\frac{13\sqrt{6}}{7560}
-\frac{\sqrt{15}}{135}\right)\mu_{\Sigma^{-}}\nonumber\\
&&+\left(-\frac{5779}{68040}-\frac{73\sqrt{2}}{5040}
+\frac{73\sqrt{3}}{7560}+\frac{73\sqrt{6}}{15120}-\frac{8\sqrt{15}}{135}\right)\mu_{\Xi^{0}}
+\left(-\frac{1}{30}-\frac{\sqrt{2}}{60}
+\frac{\sqrt{3}}{90}+\frac{\sqrt{6}}{180}\right)\mu_{\Delta^{++}}.
\label{mu10usum}\eea\end{widetext}


\begin{thebibliography}{99}
\bibitem{stern33}  R. Frisch and O. Stern, Z. Physik {\bf 85}, 4 (1933).
\bibitem{cg}  S. Coleman and S.L. Glashow, Phys. Rev. Lett. {\bf 6}, 423 (1961).
\bibitem{boss}  A. Bosshard et al., Phys. Rev. D {\bf 44}, 1962 (1991).
\bibitem{die}  H.T. Diehl et al., Phys. Rev. Lett. {\bf 67}, 804 (1991).
\bibitem{lein}  D.B. Leinweber, T. Draper and R.M. Woloshyn, Phys. Rev. D {\bf 46}, 3067 (1992).
\bibitem{sch}  F. Schlumpf, Phys. Rev. D {\bf 48}, 4478 (1993);
J. Linde and H. Snellman, Phys. Rev. D {\bf 53}, 2337 (1996).
\bibitem{decup}  S.T. Hong and G.E. Brown, Nucl. Phys. A {\bf 580}, 408 (1994).
\bibitem{but}  M.N. Butler, M.J. Savage and R.P. Springer, Phys. Rev. D {\bf 49}, 3459 (1994).
\bibitem{lee1}  F.X. Lee, Phys. Lett. B {\bf 419}, 14 (1998);
F.X. Lee, Phys. Rev. D {\bf 57}, 1801 (1998).
\bibitem{lin2}  J. Linde, T. Ohlsson and H. Snellman, Phys. Rev. D {\bf 57}, 5916 (1998).
\bibitem{kim04} G.S. Yang, H.C. Kim, M. Praszalowicz and K. Goeke, Phys. Rev. D {\bf 70}, 114002
(2004).
\bibitem{jenkins93plb} E. Jenkins, M. Luke, A.V. Manohar and M. Savage, Phys. Lett. B {\bf 302}, 482 (1993).
\bibitem{musolf01} S.J. Puglia, M.J. Ramsey-Musolf and S.L. Zhu, Phys. Rev. D {\bf 63}, 034014 (2001).
\bibitem{jenkins94plb} E. Jenkins and A.V. Manohar, Phys. Lett. B {\bf 335}, 452 (1994).
\bibitem{jenkins02prl} E. Jenkins, X. Ji and A.V. Manohar, Phys. Rev. Lett. {\bf 89}, 242001 (2002).
\bibitem{lebed04} R.F. Lebed and D.R. Martin, Phys. Rev. D {\bf 70}, 016008 (2004).
\bibitem{kim05} H.C. Kim, M. Polyakov, M. Praszalowicz, G.S. Yang and K. Goeke, Phys. Rev. D {\bf 71},
094023 (2005).
\bibitem{sample04} D.T. Spayde et al. [SAMPLE Collaboration], Phys. Lett. B {\bf 583}, 79 (2004).
\bibitem{mck89}  R.D. McKeown, Phys. Lett. B {\bf 219}, 140 (1989);
E.J. Beise and R.D. McKeown, Comm. Nucl. Part. Phys. {\bf 20}, 105
(1991); R.D. McKeown, {\it New Directions in Quantum
Chromodynamics} (AIP, Melville, New York 1999) eds. C.R. Ji, D.P.
Min.
\bibitem{happex06} A. Acha et al. [HAPPEX Collaboration], Phys. Rev. Lett. {\bf 98}, 032301 (2007).
\bibitem{mck}  R.D. McKeown, hep-ph/9607340 (1996).
\bibitem{gerry791}  G.E. Brown and M. Rho, Phys. Lett. B {\bf 82}, 177 (1979).
\bibitem{hong93} S.T. Hong and B.Y. Park, Nucl. Phys. A {\bf 561}, 525 (1993);
S.T. Hong, B.Y. Park and D.P. Min, Phys. Lett. B {\bf 414}, 229
(1997).
\bibitem{hongpr01} S.T. Hong and Y.J. Park, Phys. Rep. {\bf 358}, 143 (2002), and references therein.
\bibitem{jenkins94} R.F. Dashen, E. Jenkins and A.V. Manohar, Phys. Rev. D {\bf 49}, 4713 (1994);
R. Dashen, E. Jenkins and A.V. Manohar, Phys. Rev. D {\bf 51},
3697 (1995); J. Dai, R. Dashen, E. Jenkins and A.V. Manohar, Phys.
Rev. D {\bf 53}, 273 (1996).
\bibitem{thooft74} G. t'Hooft, Nucl. Phys. B {\bf 72}, 461 (1974).
\bibitem{witten79} E. Witten, Nucl. Phys. B {\bf 160}, 57 (1979).
\bibitem{jenkins98} R. Flores-Mendieta, E. Jenkins and A.V. Manohar, Phys. Rev. D {\bf 58}, 94028 (1998).
\bibitem{lee}  M.A.B. Beg, B.W. Lee and A. Pais, Phys. Rev. Lett. {\bf 13}, 514 (1964).
\bibitem{nrqm}  S.T. Hong and G.E. Brown, Nucl. Phys. A {\bf 564}, 491 (1993).
\end{thebibliography}
\end{document}